# Development of a 10.8-eV Tabletop Femtosecond Laser with Tunable Polarization for High-Resolution Angle-Resolved Photoemission Spectroscopy

Jisong Gao,[1,2,#] Qiaoxiao Zhao,[1,2,#] Wenbo Liu,[1,2,#] Dong Li,[1,2] Zhicheng Gao,[1,2] Yudian Zhou,[1,2] Xuegao Hu,[1,2] Zhihao Cai,[1,2] Zhilin Li,[1,2] Youguo Shi,[1,2] Peng Cheng,[1,2] Zhaojun Liu,[3] Lan Chen,[1,2*] Kehui Wu,[4,5*] Zhigang Zhao,[3*] Baojie Feng[1,2,5,6*]

[1]*Institute of Physics, Chinese Academy of Sciences, Beijing,100190, China*
[2]*State Key Laboratory of Low Dimensional Quantum Physics and Department of Physics, Tsinghua University, Beijing, 100084, China*
[3]*School of Information Science and Engineering, Shandong University, Qingdao, China*
[4]*Tsientang Institute for Advanced Study, Zhejiang, 310024, China*
[5]*Interdisciplinary Institute of Light-Element Quantum Materials and Research Center for Light-Element Advanced Materials, Peking University, Beijing, 100871, China*
[6]*Songshan Lake Materials Laboratory, Dongguan, 523808, China*

[#]These authors contributed equally to this work.
[*]Corresponding author. E-mail: lchen@iphy.ac.cn; khwu@iphy.ac.cn; zhigang@sdu.edu.cn; bjfeng@iphy.ac.cn.

## Abstract

**The development of extreme ultraviolet sources is critical for advancing angle-resolved photoemission spectroscopy (ARPES), a powerful technique for probing the electronic structure of materials. Here, we report the construction of a tabletop 10.8-eV femtosecond laser through cascaded third-harmonic generation, which operates at a repetition rate of 1 MHz and delivers a photon flux of approximately $10^{12}$ photons/s. The system achieves a high energy resolution of approximately 11.8 meV and tunable polarization. This flexibility enables detailed studies of orbital**

**and (pseudo)spin characteristics in quantum materials. We demonstrate the capabilities of this laser-ARPES system by investigating several prototypical materials, showcasing its potential for elucidating complex phenomena in quantum materials.**

I. Introduction

Angle-resolved photoemission spectroscopy (ARPES) has become a cornerstone technique for exploring the electronic structure of materials with both high momentum and energy resolution[1-5]. By directly measuring the angular and energy distributions of photoemitted electrons, ARPES provides essential insights into electronic band structures, electron-phonon coupling, many-body interactions, and quasiparticle dynamics[6,7]. The ongoing development of ARPES systems is vital for advancing quantum materials research. Two primary components of an ARPES system—photon sources and electron energy analyzers—have seen significant advancements in recent decades. Electron energy analyzers have evolved from one-dimensional to two- and three-dimensional detectors, substantially improving detection efficiency[8-10]. Meanwhile, the energy resolution has advanced from several hundred meV to sub-meV levels[11]. However, the development of advanced photon sources remains a bottleneck in further enhancing ARPES performance, as the photon source determines the energy and momentum resolution, detectable momentum range, and probing depth.

The most convenient photon source for ARPES is the gas-discharge lamp, which generates monochromatic light through electric discharge in ionized noble gases. However, the light produced is unpolarized and incoherent, making it unsuitable for studying orbital characteristics and ultrafast dynamics of quantum materials. Synchrotron radiation sources provide tunable photon energy and polarization with superior beam quality, but they have their own limitations, including long pulse durations, limited temporal coherence, and prohibitive facility costs.

To overcome these limitations, tabletop lasers have emerged as promising photon sources for high-resolution ARPES measurements[12-17]. These lasers are typically generated via nonlinear frequency conversion techniques using nonlinear crystals such as beta barium borate (BBO), KBe$_2$BO$_3$F$_2$, and so on. However, the photon energies achievable with these crystals are typically limited to around 7 eV[18], which is insufficient for accessing the entire first Brillouin zone of most materials. Recent advancements in high-harmonic generation (HHG) have enabled the production of higher photon energies, extending into the X-ray regime[19,20]. However, HHG typically generates fixed linear polarization. Although circularly polarized HHG is technically possible[21,22], the complexity of the optical setup makes it impractical for integration into ARPES systems.

Among the available lasers, photon energies near 11 eV are particularly attractive for ARPES measurements. This energy is close to the absorption edges of wide-gap insulators such as Lithium Fluoride (LiF) and Magnesium Fluoride (MgF$_2$), enabling the use of windows to maintain ultrahigh vacuum in the analysis chamber without requiring differential pumping systems. Moreover, optical components such as lenses, wave plates, and prisms can be used to manipulate these lasers, unlike higher-energy photon sources. In 2016, He et al. constructed a 10.897-eV (113.778 nm) picosecond laser based on a two-photon resonant four-wave mixing technique[23]. However, this method requires a sub-nanosecond pulse width and precise control of the central wavelength to ensure a narrow spectral width and efficient two-photon resonance. Alternatively, Berntsen et al. developed a 10.5-eV (115.2 nm) picosecond laser by frequency tripling the third harmonic of a pulsed IR laser in Xe gas[24]. Recently, similar techniques have been employed to generate femtosecond 11-eV lasers for time-resolved ARPES, as demonstrated by Lee et al.[25], Kawaguchi et al.[26], and Lu et al.[27] Despite these advances, femtosecond 11-eV lasers still face significant limitations, including poor energy resolution and fixed polarization, which limit their ability to investigate complex electronic structures, orbital characteristics, and (pseudo)spin properties[28-32].

Here, we report the development of a 10.8-eV femtosecond laser with tunable polarization. When integrated into an ARPES system, it achieves an energy resolution of approximately 11.8 meV. We demonstrate the capability of the system to probe electronic structures, orbital characteristics, and (pseudo)spin textures of quantum materials with high resolution. These advancements mark a significant step toward the development of compact, high-performance ARPES systems for investigating complex quantum phenomena.

## II. System Overview

### A. 10.8-eV femtosecond laser

To improve the energy resolution of ARPES, it is essential to use lasers with high repetition rates and low pulse energy to minimize the space charge effect and thermal effects. However, low pulse energy reduces the efficiency of nonlinear optical processes, necessitating a balance between these parameters. In our system, we choose a balanced repetition rate of 1 MHz.

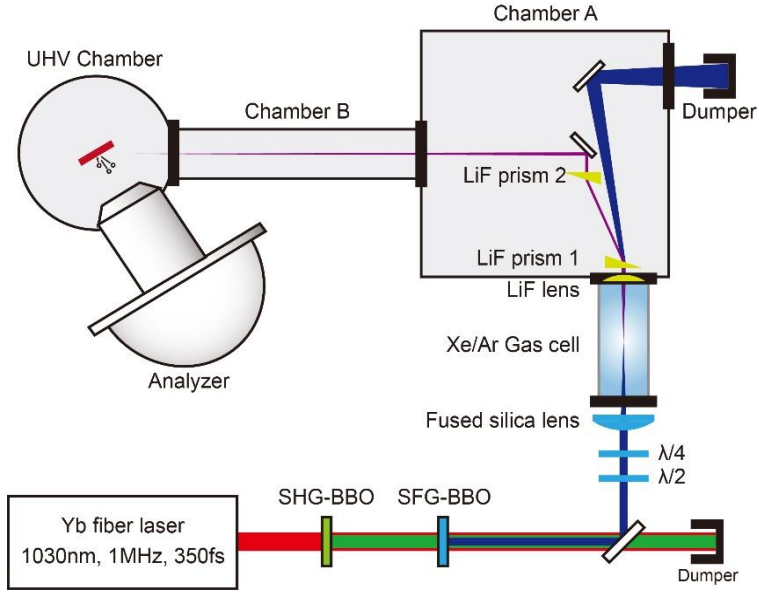

**FIG. 1.** Schematic layout of the 10.8-eV laser-ARPES system, illustrating the optical setup for laser generation and its integration with the ARPES chamber.

Our system starts with a home-built 1030 nm fiber laser system[33], which delivers 350-fs pulses centered at 1030 nm, with a repetition rate of 1 MHz and a pulse energy of 60 μJ. The entire laser system is housed in a clean room to ensure optimal performance and stability. The 1030-nm laser undergoes two stages of third-harmonic generation (THG). First, two BBO crystals convert the wavelength to 343 nm. The power of the 343-nm laser is controlled through a combination of a half-wave plate and a polarizer. The 343-nm laser is then focused into a gas-cell chamber containing a mixture of xenon (Xe) and argon (Ar) gases, where the wavelength is converted to 114.4 nm (10.8-eV)[34]. The gas cell has a length of 250 mm. The lens for the 343-nm laser is placed just before the gas cell to allow easier adjustment of the focal position. The photon flux of the 10.8-eV laser is optimized by sequentially introducing Xe and Ar gases into the gas-cell chamber, as shown in Fig. 2(a).

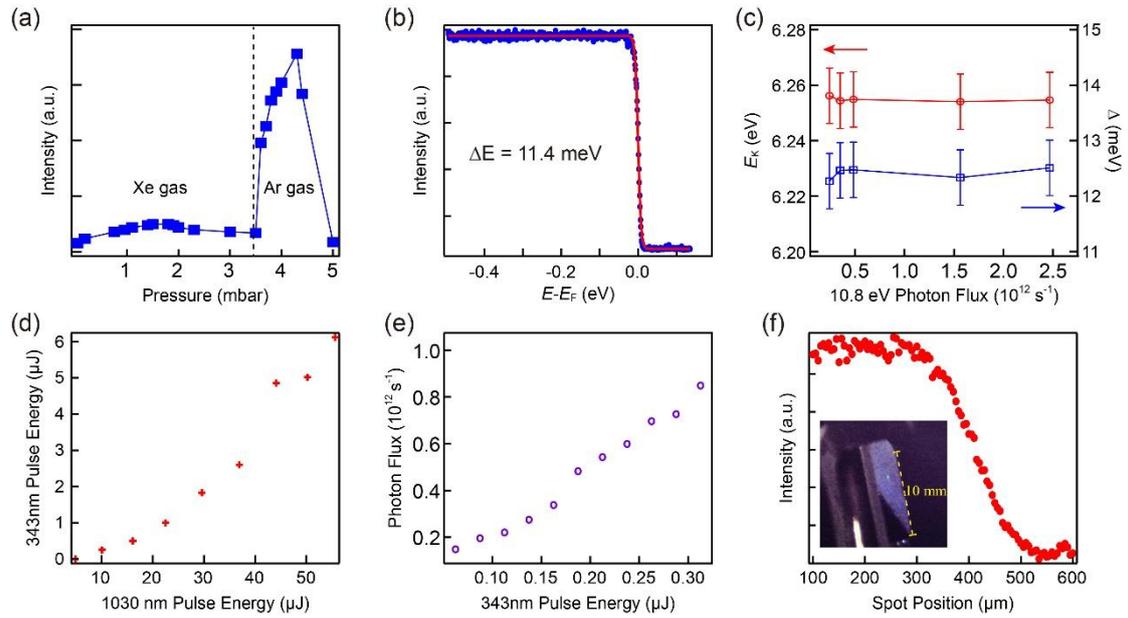

**FIG. 2.** (a) Power output of the 10.8-eV laser as a function of Xe and Ar gas pressures, measured using a calibrated photodiode. (b) Energy distribution curve of polycrystalline Au, with the red curve representing the fitting result using a Fermi-Dirac function. (c) Dependence of the Fermi level position and spectral broadening on the 10.8-eV photon flux, indicating minimal space charge effects. (d) Single-pulse energy of the 343-nm laser plotted as a function of the single-pulse energy of the 1030-nm laser after passing through two BBO crystals. (e) Photon flux of the 11-eV laser plotted as a function of the single-pulse energy of the 343-nm light. (f) Photoemission intensity as a function of Cu crystal position. The diameter of the spot is approximately 200 μm. The inset shows a picture of the 11-eV spot on a fluorescent sample.

**B. Focusing and vacuum isolation**

The gas-cell chamber is isolated from the vacuum chamber using a LiF window, which allows the 10.8-eV laser to pass through while maintaining the vacuum of the ARPES chamber. This design ensures that the vacuum level within the ARPES chamber remains stable during measurements, which is critical for preserving the long lifetime of sensitive samples. Additionally, the LiF window doubles as a lens for the 10.8-eV laser. By combining it with a 343 nm lens, the focal point of the 10.8-eV laser can be continuously adjusted to ensure a small spot size on the sample. The focal lengths for the 343-nm and 10.8-eV laser are 50 mm and 200 mm, respectively. The distance between these two lenses is approximately 280 mm; however, this distance is carefully adjusted during alignment to achieve optimal beam focusing on the sample surface, which is located approximately 2.5 meters from the final lens.

To prevent spatial distortion of the 10.8-eV laser, a pair of prisms is used to compensate for the spatial dispersion introduced when the light passes through a single prism. Each prism has an apex angle of 20°. This configuration preserves the beam quality and ensures a well-defined, high-quality beam profile. As a result, a nearly circular spot with a small diameter is achieved, providing the precision necessary for high-resolution ARPES measurements.

## C. Polarization of the 10.8 eV laser

The polarization of the 10.8-eV laser is controlled by adjusting the polarization of the 343-nm fundamental laser using half-wave and quarter-wave plates. Unlike nonlinear crystals such as BBO, which require linearly polarized fundamental light due to the presence of an optical axis, noble gases are isotropic, allowing circularly polarized fundamental light to undergo frequency tripling as well. As a result, both linearly and circularly polarized 343-nm light can be efficiently upconverted to 10.8-eV photons with the same polarization handedness. This process can be well-described within the perturbative nonlinear optics framework and is fundamentally distinct from the non-perturbative electron recollision dynamics characteristic of HHG. To maximize the photon flux, the relative pressures of Xe and Ar gases must be carefully optimized for different polarization states.

In a recently developed 10.7-eV ARPES system, a dichroic mirror (DM) was used to separate the fundamental light[26]. However, the use of DMs inherently limits the 10.8-eV photons to a fixed linear polarization. To overcome this limitation, we employ prisms to separate the 10.8-eV light from the fundamental light, enabling full polarization control while maintaining high photon efficiency.

### III. System Characterization

#### A. Energy resolution

To evaluate the energy resolution of the system, we measured the spectral width of the Fermi level of polycrystalline Au in the angle-integrated mode. The resulting energy distribution curve (EDC) is well fitted by a Fermi-Dirac function, as shown by the red line in Fig. 2(b). Our fitting analysis yields a total energy resolution of approximately 11.8 meV, which is high enough for most ARPES measurements. Within our accessible fluence range, no appreciable broadening or Fermi-level shift was observed, as show in Fig. 2(c), indicating that space charge effects are negligible under our operating conditions.

#### B. Photon flux

We measured the photon flux of the 11-eV laser as a function of the pulse energy of the fundamental 343-nm laser, as shown in Fig. 2(e). Due to the high energy conversion efficiency, we achieved a photon flux of approximately $10^{12}$ photons/s with a low pulse energy of 0.3 μJ, corresponding to an average power of 0.3 W. Although a higher photon flux can be readily obtained by increasing the power of the fundamental light, this would lead to significant space charge effects, ultimately degrading the energy resolution of the system.

#### C. Spot size

To characterize the spot size of the 10.8-eV laser, we used a copper crystal with well-defined straight edges. Figure 2(f) shows the photoemission intensity as the Cu crystal is translated across the laser spot. As the crystal moves out of the laser beam, the photoemission intensity decreases, with a transition width of approximately 200 μm.

This measurement indicates that the diameter of the laser spot is around 200 μm. The inset of Fig. 2(f) shows an image of the 11-eV laser spot, revealing its nearly circular shape.

## IV. Performance

### A. Linear dichroism ARPES

To characterize the performance of linear dichroism ARPES measurements, we investigated the Bi-√3/Ag(111) surface and Au(111). Figures 3(a) and 3(b) show ARPES spectra along the $\overline{M\Gamma M}$ direction of Ag(111) obtained using *p*- and *s*-polarized light, respectively. As expected, the characteristic Rashba-split bands near the Γ point are clearly visible with *p*-polarized light but are barely detectable with *s*-polarized light. This linear dichroism signal originates from the orbital character of these bands, consistent with previous studies[35].

Figures 3(c) and 3(d) present the ARPES spectra of the Shockley surface states of Au(111). The Rashba splitting is distinctly resolved, demonstrating the high energy and momentum resolution of our system. Similar to the Bi/Ag(111) system, the surface states of Au(111) exhibit a strong response to *p*-polarized light while being nearly invisible with s-polarized light, in agreement with previous reports[36].

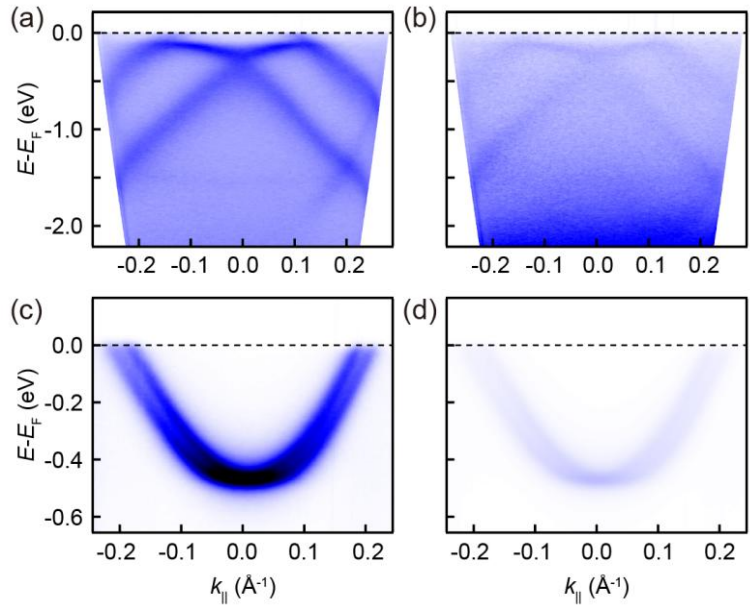

**FIG. 3.** (a,b) ARPES spectra of Bi/Ag(111) measured with *p*- and *s*-polarized light,

respectively. (c,d) ARPES spectra of the surface states of Au(111) measured with *p*- and *s*-polarized light, respectively.

## B. Circular dichroism (CD) ARPES

Figures 4(a) and 4(b) present the ARPES spectra of the Shockley surface states of Au(111) obtained with left- and right-circularly polarized light. The intensity distribution exhibits an inverse response between the left and right sides of the surface states, resulting in a pronounced CD signal, as shown in Fig. 4(c). The CD effect arises from the contribution of the *p*-to-*d* photoemission channel to the total photoemission intensity rather than from the spin texture[36].

We further performed CD-ARPES measurements on the topological insulator $Bi_2Se_3$, which hosts spin-polarized surface states protected by time-reversal symmetry[37]. Previous CD-ARPES studies have reported strong dichroic signals associated with the spin texture of topological surface states[38]. Consistent with these findings, our measurements also reveal a strong CD signal in the surface states of $Bi_2Se_3$, as shown in Figs. 4(d)-4(f). Therefore, our ARPES system has the capability for spin-sensitive measurements.

The differential CD maps shown in Figs. 4c and 4f offer an indirect yet informative approach to estimating the degree of circular polarization (DOCP) of the 10.8-eV laser. The Rashba-split surface states of Au(111) and the topological surface states of $Bi_2Se_3$ exhibit nearly symmetric band dispersions and well-understood dichroic responses under circularly polarized excitation. By analyzing the absolute intensity ratio between the left and right branches in these differential maps, we estimate a lower bound for the DOCP. In our measurements, this ratio consistently exceeds 0.9, indicating that the DOCP of our laser source surpasses 90% under optimized gas mixing and polarization control conditions.

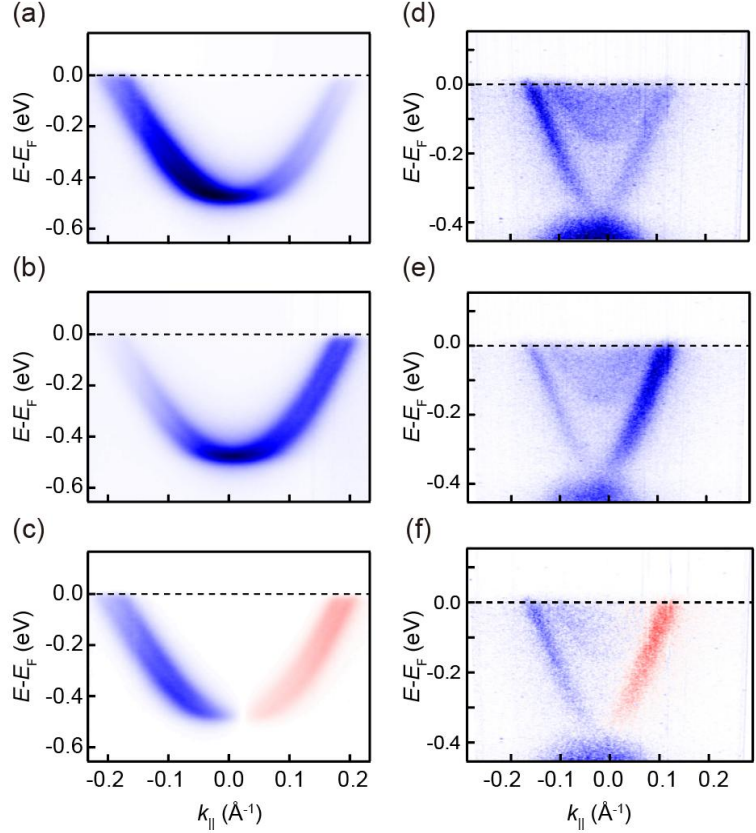

**FIG. 4.** (a,b) ARPES spectra of Au(111) measured with left- and right-circularly polarized light. (c) Differential image showing the circular dichroism signal obtained by subtracting (b) from (a). (d-f) Same as (a-c), but for the surface states of $Bi_2Se_3$.

## V. Summary and Outlook

In summary, we have developed a tabletop 10.8-eV femtosecond laser using cascaded third-harmonic generation and integrated it into an ARPES system. The laser operates at a repetition rate of 1 MHz with an energy resolution of approximately 11.8 meV. A key advantage of this system is its tunable polarization, enabling both linear and circular polarization states, which facilitates the study of orbital and (pseudo)spin-dependent properties in quantum materials. The system design also incorporates a LiF window for vacuum isolation, eliminating the need for differential pumping while maintaining ultrahigh vacuum conditions. The performance of the system was validated through ARPES measurements of prototypical materials, including Bi/Ag(111), Au(111), and $Bi_2Se_3$, revealing strong linear and circular dichroism signals.

This high-resolution laser-ARPES system opens new avenues for exploring emergent phenomena in quantum materials, including electron correlation effects, topological phases, and ultrafast dynamics. Future advancements could focus on enhancing the energy resolution, increasing the photon flux, and extending the photon energy beyond 10.8 eV to access a broader momentum range. Furthermore, the femtosecond pulse width enables the integration of pump-probe techniques for time-resolved ARPES, allowing direct investigations of non-equilibrium dynamics in complex materials.

**Acknowledgements**

This work was supported by the National Key R&D Program of China (Grants No. 2024YFA1408700 and 2024YFA1408400), the National Natural Science Foundation of China (Grants No. W2411004, 12374197, T2325028, and 12134019), the Beijing Natural Science Foundation (Grant No. JQ23001), and the CAS Project for Young Scientists in Basic Research (Grants No. YSBR-047).